\documentstyle[prb,aps,psfig,twocolumn]{revtex}
\begin{document} 
\title{Role of bound pairs in the optical properties of highly excited
semiconductors: a self consistent ladder approximation approach.}

\twocolumn[\hsize\textwidth\columnwidth\hsize\csname@twocolumnfalse\endcsname
\author{C. Piermarocchi} \address{Department of Physics, University of
California San Diego, La Jolla, CA 92093-0319} \author{F. Tassone}
\address{Scuola Internazionale Superiore di Studi Avanzati, via Beirut
4, I-34014 Trieste, Italy} \date{\today} \maketitle 

\begin{abstract}
Presence of bound pairs (excitons) in a low-temperature electron-hole
plasma is accounted for by including correlation between fermions at
the ladder level. Using a simplified one-dimensional model with
on-site Coulomb interaction, we calculate the one-particle
self-energies, chemical potential, and optical response.  The results
are compared to those obtained in the Born approximation, which does
not account for bound pairs.  In the self-consistent ladder
approximation the self-energy and spectral function show a
characteristic correlation peak at the exciton energy for low
temperature and density. In this regime the Born approximation
overestimates the chemical potential.  Provided the appropriate vertex
correction in the interaction with the photon is included, both ladder
and Born approximations reproduce the excitonic and free pair optical
absorption at low density, and the disappearance of the exciton
absorption peak at larger density.  However, lineshapes and energy
shifts with density of the absorption and photoluminescence peaks are
drastically different.  In particular, the photoluminescence emission
peak is much more stable in the ladder approximation. At low
temperature and density a sizeable optical gain is produced in both
approximations just below the excitonic peak, however this gain shows
unphysical features in the Born approximation. We conclude that at low
density and temperature it is fundamental to take into account the
existence of bound pairs in the electron-hole plasma for the
calculation of its optical and thermodynamic properties.  Other
approximations that fail to do so are intrinsically unphysical in this
regime, and for example are not suitable to address the problem of
excitonic lasing.

\end{abstract} \pacs{}

]


\section{Introduction} 

The binding of a gas of oppositely charged fermions into a bosonic gas
of bound pairs has a fundamental interest in condensed matter physics,
in problems ranging from the state of hydrogen to the nature of the
electron-hole plasma in Si and Ge. This pairing takes its origin from
the Coulomb correlation between the charged fermions, and strongly
affects the statistical and thermodynamic properties of these systems.
In semiconductor physics, the role of the Coulomb correlation and
pairing in the description of the electron-hole plasma in Si and Ge
has been extensively investigated. Refined descriptions of both the
ground state and the thermodynamic properties of this system have been
developed in the past thirty years. In particular, an excitonic
insulator ground state was first proposed and then shown to be less
stable than a simpler electron-hole liquid in Si and Ge, due to
band-structure effects.\cite{rice77} Thermodynamics of the
electron-hole gas in simpler crystals was also addressed by
Haug\cite{haug76}, and Zhu {\it et al.}\cite{zhu96} aiming at deriving
a complete phase diagram from variational approaches.

In this paper we will deal with direct gap semiconductors and we will
discuss the thermodynamics of the photoexcited electron and holes at
quasi-thermal equilibrium. Our main concern is to determine how the
inclusion of pairing effects in the theoretical description, as done
for the systems mentioned above, can affect predictions on the shape
of optical spectra. In the optical properties of semiconductors, the
effect of the Coulomb electron-hole correlation comes into play to
explain the characteristic excitonic absorption and emission. These
excitonic features can be described by including vertex corrections in
the interaction of the electron-hole pair with the
photon.\cite{elliott57} In particular, excitonic absorption is found
in an empty crystal, when a background electron-hole plasma is absent.
However, when carriers are electrically or optically injected, they
reach quasi-thermal equilibrium through scattering, and a background
electron-hole plasma is eventually formed. In this typical situation
the absorption is modified, and spontaneous emission
(photoluminescence) also takes place. Absorption and emission
give valuable information about the state of the plasma, but a theory
describing both the thermodynamics and optical properties of the
plasma is required in order to extract this information.
Semiconductor Bloch equations, and their evolutions, are certainly
among the most known and used theories for this
purpose.\cite{haugkoch} In these theories, the electron-hole plasma
has been originally treated at the Hartree-Fock level,\cite{haug84}
and eventually screening of the Coulomb interaction was also included
at various levels of approximation.\cite{haugkoch} However, none of
these refinements accounts for the existence of bound excitons in the
plasma. This is certainly a weak point at low temperatures, when
simple thermodynamic arguments show that condensation of electron-hole
gas into a gas of bound-electron hole pairs is favorable in a wide
range of densities.

In the following sections we will show the relevant changes in the optical
properties of the semiconductor when the existence of excitons at low
temperatures is taken into account. This is clearly a leap beyond the
standard description of the interaction of light with the
semiconductor as described above.  Our purpose is qualitative, but we
can still gain solid understanding of the underlying physics. In our
opinion at this stage of development it is not wise to aim at
improving also quantitatively existing theories which are already far
too involved.  We keep the problem simple enough so as to always control
and understand the resulting physics. We thus consider an accurately
solvable model, where the system is one-dimensional, Coulomb
interaction is on-site, and electron and holes are spin-polarized.
Excitonic correlation is also expected to be stronger in one dimension
than higher dimensions. Moreover, for the single-pair problem, exactly
a single bound state is obtained with the on-site interaction.  In
our description of the electron-hole plasma, we either include carrier
scattering at the Born level, which does not describe bound pairs in
the plasma, or we include enough correlation so as to describe the
pairing into bound states. For this purpose, ladder diagrams in the
electron-hole scattering kernel\cite{kraeft86,zimmermann88,pereira98}
are included, and the electron and hole self energies calculated with
it.\cite{tassone99} The resulting Dyson equations for the Green
functions are solved self-consistently. The approximation is therefore
called self-consistent ladder approximation (SCLA). Screening is
expected to be weak in one dimension,\cite{benner91} due to reduced
screening phase-space, and we neglect it altogether.  We remark that
in the SCLA we calculate the single particle propagators and
self-energies only, thus in this sense, the theory remains purely
fermionic. We do not map the model onto bound and un-bound states at
any time. However, it is possible to show analytically that at low
temperatures and densities, a bosonic gas of excitons is effectively
described, and that self-consistency effectively introduces scattering
between these excitons. When this analogy is extended to larger
densities and temperatures, where bound and free carriers are expected
to coexist, we understand that self-consistency also accounts for
multiple exciton-free carrier, and free carrier-free carrier
scattering. Thus, a large amount of correlation, well beyond that
described in simpler Born approximations is included in the SCLA.  We
remark that we also assume quasi-thermal equilibrium in the electron-hole
gas, which is well justified as the characteristic radiative
recombination (several hundreds of picoseconds) is much slower than
the typical scattering time in the density range considered. The
theory is however written so as to be readily extended to the
non-equilibrium situation. We finally show how to correctly calculate
photon absorption and emission (photoluminescence or PL) in a
conserving sense, i.e. respecting f-sum rules. We only neglect
polaritonic effects, which are very weak in one
dimension.\cite{tassone.93}

Within the considered models, we highlight important trends in optical
and thermodynamical properties.  First, Born approximations
overestimate the chemical potential at low temperatures when a
consistent fraction of pairs is bound into excitons. Second, energetic
stability of the excitonic emission peak as a function of density is
better described in the SCLA, even well beyond disappearance of
excitonic absorption peak at large excitation densities. Third,
the large excitonic {\em gain} at densities just below those
where excitonic absorption disappears is clearly incorrectly described
in simpler Born approximations.  As these trends have a clear physical
origin, they are expected to hold qualitatively even when more
realistic interaction potentials (eventually including screening) are
considered. We conclude that these results do have important
experimental and theoretical relevance, signaling the necessity of an
adequate description of the excited semiconductor in the
low-temperature region. For example, current description of the
electron-hole plasma in the semiconductor Bloch equations is clearly
insufficient to address excitonic (inversion-less) lasing.

The paper is organized as follows. In Sec. \ref{sec:method} we
introduce and explain how to implement the self consistent ladder
approximation, and the Markov-Born approximation and self-consistent
Born approximation.  Single particle properties are analyzed in
Sec. \ref{sec:single}, where we show how excitonic correlation appears
in the electron and hole spectral functions.  Chemical potentials and
self-energies are then compared. In Sec. \ref{sec:excpot} we
analytically show how the SCLA at low temperature and density
describes an interacting exciton gas. Optical properties of a
semiconductor quantum wire are calculated in Sec. \ref{sec:optical}
for the three approximations. Absorption and emission as functions of
the density are also presented, and discussed in this
section. Conclusions are drawn is Sec. \ref{sec:conclusions}.

\section{Born and self consistent ladder approximation}
\label{sec:method}

\begin{figure}
\centerline{\psfig{figure=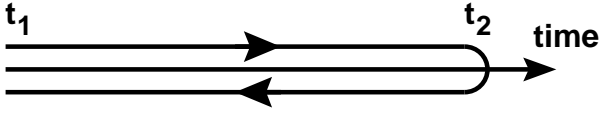,width=8truecm}}
\vspace{1cm}
\caption{The Keldysh contour. Upper branch, normal time ordering,
lower branch, reversed time ordering.}
\label{fig:keldysh}
\end{figure}

In this section we introduce the SCLA, the simpler Markov-Born
approximation (MBA) and self-consistent Born approximation (SCBA). We
assume thermal equilibrium in the electron-hole plasma, but, in order
to avoid analytic continuation problems, we do not resort to the
Matsubara Green function technique,\cite{mahan81} which becomes
numerically delicate when the spectral function shows a large number
of poles. Instead, we work directly in the real time/frequency
space. This has the additional advantage that the approach can be
readily extended to non-equilibrium.\cite{danielewicz} We thus
consider four single particle Green's Functions:\cite{haug96} the
retarded, advanced, lesser and greater Green's functions defined as
\begin{mathletters}
\label{eq:greens}
\begin{eqnarray} 
G^+(1,2)&=&-i \theta(t_1-t_2) \langle T [c(1) c^\dagger(2)]_+ \rangle,
\label{eq:g+} \\ G^-(1,2)&=&i \theta(t_2-t_1) \langle T [c(1)
c^\dagger(2)]_+ \rangle, \label{eq:g-}\\ G^<(1,2)&=&i \langle
c^{\dagger}(2) c(1) \rangle, \label{eq:g<}\\ G^>(1,2)&=&-i \langle
c(1) c^{\dagger}(2)\rangle . \label{eq:g>}
\end{eqnarray}
\end{mathletters}
Here $T$ is the time ordering operator, $1=(x_1,t_1,\sigma_1)$, where
$x_1$ is the position, $t_1$ the time, and $\sigma_1$ is the spin
index. $c(1)$ is the electron annihilation operator and $[~ ]_+$ is
the anticommutator.  Similar expressions for the hole Green's
functions hold, with $ d(1)$ and $ d^{\dagger}(1)$ the hole
annihilation and creation respectively.  Only two of the four
functions above are independent, and in particular, we will consider
the retarded and lesser (or correlation) functions. Formally, a single
Green's function $G(x_1,\tau_1,x_2,\tau_2)$, may be introduced for
compactness of notation, with $\tau_1$ and $\tau_2$ defined on the
Keldysh contour, having the two branches shown in
Fig. \ref{fig:keldysh}. Following simple rules, any equation with
times defined on the Keldysh contour can be written back in a set of
equations defined on the ordinary time axis for the above four Green
functions (see for instance Ref. \onlinecite{haug96}).
The statistical averages are done in the macro-canonical ensemble:
$\langle \dots \rangle={\rm Tr}\{\rho \dots\}$ with $\rho=
e^{-\beta(H-\mu_{e} N_e -\mu_{h} N_h)}/{\rm Tr} \{e^{-\beta(H-\mu_{e}
N_e-\mu_{h} N_h)}\}$.  Here $\beta=1/T$ and $H$ is the total
Hamiltonian of the interacting electron-hole system.  In principle,
electrons and holes have independent chemical potentials, and
densities. However, we are interested in the description of a laser
excited semiconductor where the density of electrons and holes is the
same, and charge is balanced.  We will also assume that the
temperature of the two components of the interacting gas is the
same. In fact, after excitation, thermalization of carriers is mainly
driven by exchange of energy with the phonon thermal bath which, at
equilibrium, leads to $T_e = T_h = T $.  Moreover, for simplicity we
consider the same electron and hole masses, giving $\mu_e=\mu_h=\mu$
and the same Green functions and self-energies. Extension of the
theory to different masses is straightforward. Finally, the Green
functions above depend only on the time difference $t_2-t_1$ as the
system is stationary, and on the relative distance $x_2-x_1$ as the
system is homogeneous.

 In this thermal regime, the retarded and lesser Green functions
defined above are also related by the Kubo Martin Schwinger relations
\cite{kadanoff62}
\begin{mathletters}
\label{eq:KMSfermi}
\begin{eqnarray}
G^<(k,\omega)=-2 \Im [G^+(k,\omega)] f(\omega-\mu)\\ G^>(k,\omega)=2
\Im [G^+(k,\omega)](1-f(\omega-\mu)),
\end{eqnarray}
\end{mathletters}
where we considered the Fourier transforms with respect to the
relative time $t_1-t_2$, and with respect to the relative position
$x_1-x_2$. The function $f(\omega)=[exp(\omega/T)+1]^{-1}$ is the
Fermi function. In the stationary case
$G^{+}(k,\omega,\sigma)={G^{-}}^*(k,\omega,\sigma)$ contain the
information on the spectral properties of the quasiparticles, given by
the one particle spectral function $A(k,\omega,\sigma)=-2 \Im
[G^+(k,\omega,\sigma)]$.  The correlation functions instead contain
information on the quasi particle occupation number
$G^<(k,t_1=t_2=t,\sigma)=N(k,t,\sigma)$, and
$G^>(k,t_1=t_2=t,\sigma)=1-N(k,t,\sigma)$, where $N(k,t\sigma)$ are
the occupation numbers.

The on-site Coulomb interaction reads:
\begin{equation}
V(x) = \pm a \delta(x)~,
\label{eq:delta}
\end{equation} 
the potential is repulsive for electron-electron and hole-hole, and
attractive for electron-hole interaction. An attractive delta-like
potential between the electrons and the holes gives one bound state
only for the pair problem, and $a$ can be chosen such to reproduce the
exciton binding energy measured in the limit of low density.  The
constant $a$ has the dimensions of $energy \times length$, the exciton
Bohr radius is given by $a_B=\hbar^2/(m a)$ and the exciton binding
energy is $E_b=m a^2/2 \hbar^2$ where $m$ is the reduced mass of the
electron-hole system. We will use the units $\hbar=a=m=1$ throughout
the paper. In these units $a_B=1$ and $E_b=0.5$. We remind that
typical values for semiconductor quantum wires are $a_B=10^{-6}$ cm
and $E_b=10$ meV.\cite{optwires} With the above effective potential of
Eq. (\ref{eq:delta}) the interaction Hamiltonian is
$$H_C = \frac{1}{2}\sum_{1} c^\dagger (1)c^\dagger(1')c(1')c(1)+$$
$$+ \frac{1}{2}\sum_{1} d^\dagger (1)d^\dagger(1')d(1')d(1)+$$
\begin{equation}
-\sum_{1}
c^\dagger(1)d^\dagger(1')
c(1)d(1')~.\label{eq:H}
\end{equation} 
Here $1'=(x_1,t_1,\sigma'\ne \sigma)$ for the electron-electron and
hole hole interaction, while $1'=(x_1,t_1,\sigma')$ for the
electron-hole interaction, as due to the fermionic nature of the
carriers, only interaction with opposite spin is allowed in the
intraband term for a contact potential.  In the following, we assume a
spin-polarized system of electrons and holes, so that we keep only the last
term describing the electron-hole interaction in the Hamiltonian
(\ref{eq:H}). Generalization to both spins is straightforward.

\begin{figure}
\centerline{\psfig{figure=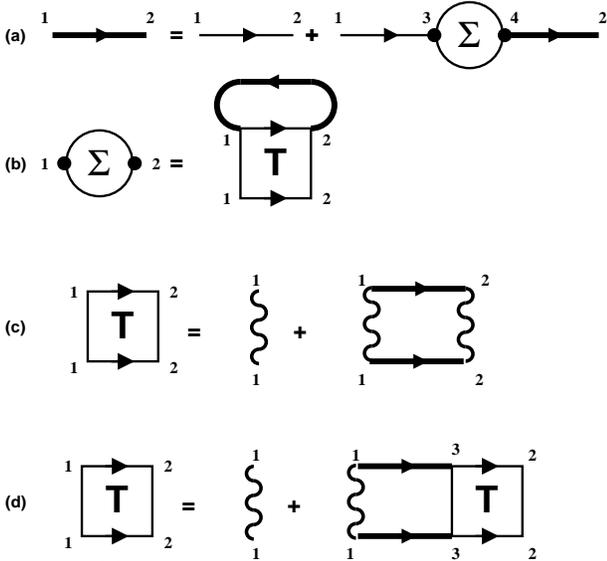,width=8truecm}}
\caption{(a) The Dyson equation, thick lines are the dressed
one-particle Green function, thin lines the bare ones. (b) The
self-energy with the scattering kernel $T(1;2)$, for the on-site
Coulomb potential. (c) The scattering kernel $T(1;2)$ for the Born
approximations. (d) The scattering kernel $T(1;2)$ in the ladder
approximation, given by the Bethe-Salpeter equation.}
\label{fig:diagrams}
\end{figure}

The Dyson equation for the single particle $G(1,2)$ propagators read
\begin{equation}
G(1,2)=G_0(1,2)+\int
d\bar 3~d\bar 4 G_0(1,\bar 3)\Sigma(\bar 3,\bar 4)G(\bar 4,2)~, $$
\label{eq:dyson}
\end{equation}
where time integrations are over the Keldysh contour.  $G_0$ is the
free propagator. The functional dependence of the self energy $\Sigma$
on the single particle Green's functions determines the degree of
approximation. In particular, it can be expressed through an
electron-hole scattering kernel $T(1,2;1',2')$:
\begin{equation}
\Sigma(1,2)= i \int d\bar 1'~d\bar 2'~T(1,\bar 1';2,\bar 2') G(\bar
2',\bar 1') \label{eq:sigma}
\end{equation}
In the Born approximation,
$$T(1,1';2,2')=V(1,1')+$$
\begin{equation} +
 V(1,1')i G(1,2)G(1',2') V(2,2'), $$
\label{eq:tborn}
\end{equation}
where $V(1,1')= V(x_1-x_{1'}) \delta(t_1,t_{1'})$, and $\delta(\tau,\tau')$
is the Dirac delta function extended on the Keldysh contour.  The
self-energy $\Sigma(k,\omega)$ can be calculated at the single
particle pole $\omega = k^2/4 $. In this case we are neglecting memory
terms in the scattering and thus using a Markov approximation; this
corresponds to the MBA. In the SCBA, the same scattering kernel of
Eq. \ref{eq:tborn} is used, but the full energy dependent structure of
the resulting self-energy given in Eq. (\ref{eq:sigma}) is used in the
Dyson equations Eq. (\ref{eq:dyson}).  The problem is then solved
self-consistently, as explained later. In this case, memory effects,
beyond the Markov approximation, are also included.  In the SCLA, the
kernel $T(1,1';2,2')$ is the solution of the Bethe Salpeter equations
in the ladder approximation, which for a generic potential reads
$$T(1,1';2,2')= V(1,1') +$$
\begin{equation} 
+\int d\bar 3~d\bar 3'~V(1,1')i G(1,\bar 3)G(1',\bar 3') T(\bar 3,\bar 3';2,2')
\label{eq:bethesalpetereh}
\end{equation} 
For a $\delta$-like potential, $V(1,1')=-\delta(1,1')$ and $T$ depends
only on $2-1$. The reduced equation reads:
\begin{equation} 
T(1;2)= -1 - i \int d{\bar 3}~G(1,\bar 3)G(1,\bar 3) T (\bar 3;2)
\label{eq:bethesalpetereh2}
\end{equation}
 The Eqs. (\ref{eq:dyson}), (\ref{eq:sigma}) and
(\ref{eq:bethesalpetereh}) are schematically represented in
Fig. \ref{fig:diagrams} for both the SCBA and the SCLA.

In the self-consistent approximation we solve self-consistently
Eqs.(\ref{eq:dyson}) and (\ref{eq:sigma}) with the scattering kernels
given by Eq. (\ref{eq:tborn}) or (\ref{eq:bethesalpetereh}) in the
Born or SCLA case respectively.  We explain in details how numerical
self consistency for a fixed temperature and chemical potential $\mu$
has been implemented.

(i) We start with ${G^+}(k,\omega)$ of the form ${G^+}_{0}=(\omega-
k^2/4+i \gamma)^{-1}$, and ${G^<}_0(k,\omega)=-2 \Im
[{G^+}_{0}(k,\omega)]~f(\omega-\mu)$.

(ii) In order to take advantage of the local character of the
electron-hole interaction, we use the Fourier transform of the Green
functions of Eq. (\ref{eq:greens}), $G(1,2)$, which actually have only
a spatial dependence on $r=x_1-x_2$ as the system is homogeneous.  We
calculate the free pair propagator defined as
\begin{equation}\label{eq:hr}
 H(1;2)= i G(1,2) G(1,2)
\end{equation} 
The retarded and correlation parts are calculated passing from the
Keldysh contours to the usual time axis:
\begin{equation}\label{eq:hr<}
 H^<(r,t)= i G^<(r,t) G^<(r,t),
\end{equation}\begin{equation}
 H^+(r,t)= i G^+(r,t) G^+(r,t)+
\label{eq:hr+} 2 i G^+(r,t) G^<(r,t).
\end{equation}
The dependence on the relative distance has been explicitly
shown. The dependence on the relative propagation time $t=t_2-t_1$
stems from the stationary condition. The pair propagator is then
Fourier transformed to $(k,w)$ space.

(iii.a) For the Born approximations, we use Eq. (\ref{eq:tborn})
\begin{mathletters}
\begin{eqnarray}
T^+(q,\omega)=-1+H^+(q,\omega)\\ T^<(q,\omega)=H^<(q,\omega)
\end{eqnarray}
 \end{mathletters}

(iii.b) For the SCLA case, Eq. (\ref{eq:bethesalpetereh2}) is readily
solved as:
\begin{mathletters}
\begin{eqnarray}\label{eq:t+w}
T^+(q,\omega)=-(1+H^+(q,\omega))^{-1}~,\\
T^<(q,\omega)=|T^+(q,\omega)|^2 H^<(q,\omega)~.
\end{eqnarray}
\end{mathletters}

(iv) The electron self-energy is then calculated in real space as in
Eq. (\ref{eq:sigma}). The retarded and correlation functions, defined
on the usual time axis, read:
$$
\Sigma^<(r,t)= i T^<(r,t)G^>(-r,-t)=-i T^<(r,t)[G^>(r,t)]^*,$$
$$
\Sigma^+(r,t)= - i T^<(r,t) G^-(-r,-t) +i T^+(r,t) G^<(-r,-t)=
$$
$$= - i T^<(r,t) [G^+(r,t)]^*+ i T^+(r,t) [G^<(r,t)]^*.$$

(v) The new electron Green's functions are then calculated as
\begin{mathletters}
$$
G^+(k,\omega)=(\omega- k^2/4+\Sigma^+(k,\omega)+i\gamma)^{-1},$$$$
G^<(k,\omega)=-2 \Im [G^+(k,\omega)]~f(\omega-\mu).$$
\end{mathletters}

The procedure is repeated through step (ii) until self-consistence is
reached. Fast Fourier transforms are performed on a finite grid of
16384$\times$128 points for the frequency and wavevector domains
respectively.  The frequency and wavevector domain $(k,w)$ is
(-20,20)X(-3,3).  Due to the finite k-range, we obtain $E_b=0.4$. The
external $\gamma>0$ in (v) is adiabatically switched off during 
self consistency. In this way the imaginary part of the final $G$ is
provided by the interaction only. For the non-self-consistent Born
approximation (or Markov-Born), we use the frequency independent,
on-pole $\Sigma^+_{k}(\omega=k^2/4)$ in the Green function, and stop
the procedure at step (v).

All of the considered approximations are conserving in the sense of
Kadanoff and Baym:\cite{baym61} the total density, total momentum,
total energy and total angular momentum of the system are
conserved. This is a fundamental property of any approximation for the
self-energy, otherwise unphysical results may result.  We
will come back to this point again in the calculation of the optical
response of the system in the various approximations.

  The polarized electron-hole gas with a contact potential interaction
that we consider here can be mapped onto a single band Hubbard model
with spin 1/2 and attractive interaction, in the limit of infinite
width of the band, and infinite on-site interaction so as to produce
finite $E_b$ (continuum limit). For the Hubbard model with repulsive
interaction, the self-consistent ladder approximation is known as
fluctuation exchange approximation (FLEX).\cite{flex} Even though in
this repulsive case the opening of an Hubbard gap is not reproduced,
in the attractive case the gap is of different nature and is
notoriously well described within this approximation.\cite{kagan98}

\section{Single Particle properties}
\label{sec:single}

\begin{figure} 
\centerline{\psfig{file=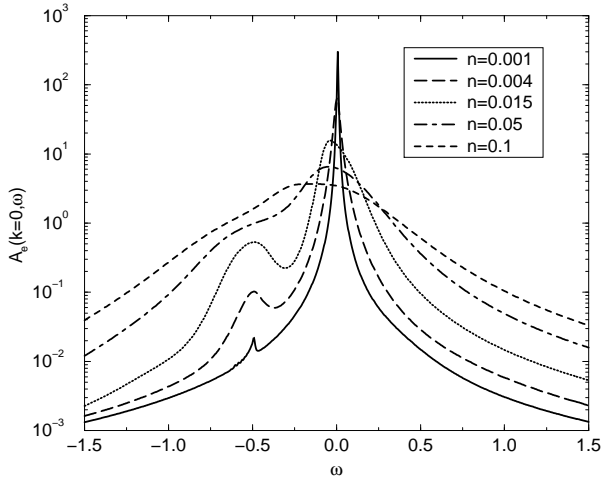,width= 9 cm}}  
\caption{Electron spectral function at $k=0$ as a function of the
total carrier density. $T=0.1$.}
\label{fig:spectral1}
\end{figure}
We plot in Fig. \ref{fig:spectral1} the electron spectral function at
$k=0$ for different densities, obtained in the SCLA. The temperature
$T=0.1\ll E_b=0.5$. For this particular figure we used the k-range
(-6,6) in the numerical solution. For $n\sim 0.001$ the electron
spectral function has a very narrow Lorentzian shape.  For larger
densities, a satellite structure appears in the low energy side of the
spectral function. This structure is located at the exciton binding
energy, below the main peak, and accounts for the correlation of the
electron at $k_e=0$ bound with holes in other $k_h$ states. Excitons
with all values of the center of mass wavevectors are involved in this
peak,\cite{zimmermann88} as we will also show in the next section.
This correlation structure is of course not present in the Born
approximations, which maintain their single quasi-particle peak
structure at any density. As the density rises, the relative weight of
the satellite structure in the SCLA spectral function with respect to
the main quasi-particle peak increases. Moreover, both structures
become broader. For densities $n>0.1$, it becomes difficult to
distinguish between the excitonic and the main peak, and a single
broad, red-shifted quasi-particle structure appears. The merging of
the excitonic peak with the single quasi-particle peak can be
interpreted as a Mott transition. We remark however that it is
difficult to give a rigorous estimate of the Mott density, because
broadening is gradual and a sharp threshold is not evident.

\begin{figure} 
\centerline{\psfig{file=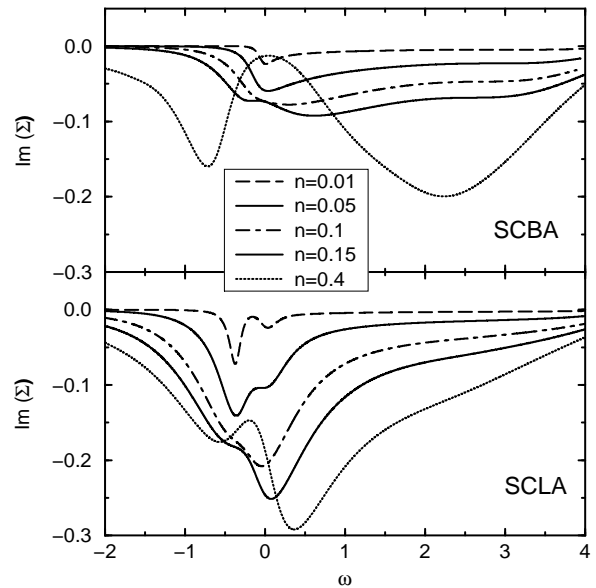,width=10 cm}}
\caption{Self energy $\Sigma (k=0,\omega) $ for $T=0.1$ and different
densities given in the Figure, in the self consistent Born
approximation (SCBA) and self-consistent ladder approximation
(SCLA).}
\label{fig:self}
\end{figure}
In Fig. \ref{fig:self} we show the imaginary part of the electron self
energy for different densities. We will also refer in the following to
this quantity as (energy dependent) broadening. In the SCLA
$\Im(\Sigma)$ shows two peaks corresponding to dephasing experienced
by an unbound or a bound state propagated in the system. Both peaks
increase with density. At low density, $n=0.01$, dephasing at the
exciton energy dominates, while at $n=0.05$ both peaks become
comparable.  In the SCBA, the peak of $\Im(\Sigma)$ at $-E_b$ is
obviously absent, as propagation of bound pairs is not allowed in the
theory. Instead at low density $n=0.01$, the broadening at the main
quasi-particle energy ($\omega=0$) is similar in the SCLA and
SCBA. Indeed, few excitons are expected in the plasma, and scattering
mainly originates from free carriers.  An heuristic understanding may
be obtained with the action mass law relating the concentration of
different chemical species in a reaction. In our case, electron+hole
$\leftrightarrow$ exciton, and the law states ${n_c}^2/n_X=n^*(T)$,
where $n_c$ and $n_X$ represent the density of unbound carriers and
excitons respectively. $n^{*}(T)$ is a cross-over density which
depends only on the temperature, and at T=0.1, $n^{*}\sim 0.005$ . For
$n < n^*$, $n_X \ll n$. Therefore, a small density of excitons is
expected at $n=0.01\sim n^*$. At this density the exciton-free carrier
contribution to free carrier broadening is weaker than the free
carrier-free carrier contribution.  At larger density, bound excitons
in the plasma become dominant, and broadening at the quasi-particle
peak becomes much larger in the SCLA than in SCBA. This trend is
observed up to $n=0.15$ indicating relevance of correlated states in
the plasma even at this large density. However, the heuristic
interpretation of the broadening based on the action mass law given
above becomes meaningless, as it is not possible to distinguish
between two chemical species anymore.  At the highest considered
density $n=0.4$, we clearly observe a dip to very small broadening in
the self-energy for SCBA case. This dip is exactly at the Fermi energy
and accounts for the blocking of the scattering at the Fermi level. In
fact, only at this large density the Fermi gas becomes degenerate
(Fermi energy much larger than T). In the SCLA, the broadening never
vanishes, but is only partially reduced at the Fermi level. This is
indicative of a more complicated structure of the electron-hole plasma
and of the broadening process, where Pauli blocking looses much of its
effectiveness, and is suggestive of a non-Fermi liquid behavior.  We
finally noticed that $\Re(\Sigma)$ is comparable to $\Im(\Sigma)$ at
any density in the SCLA, and thus remarked that the excitonic
satellite peak in the spectral function does not correspond to a zero
of $\omega-\Re(\Sigma(\omega))$. Indeed, we do not expect appearance
of another simple quasi-particle in the plasma. For $\omega\sim -E_b$,
and for $E_b>>|\Im(\Sigma)|$, we have $ A_{k=0}\sim
|\Im(\Sigma)|/|E_b|^2$. Thus, the correlation satellite in the
spectral function follows the peak of $\Im(\Sigma)$ at $\omega=-E_b$.

\begin{figure} 
\centerline{\psfig{file=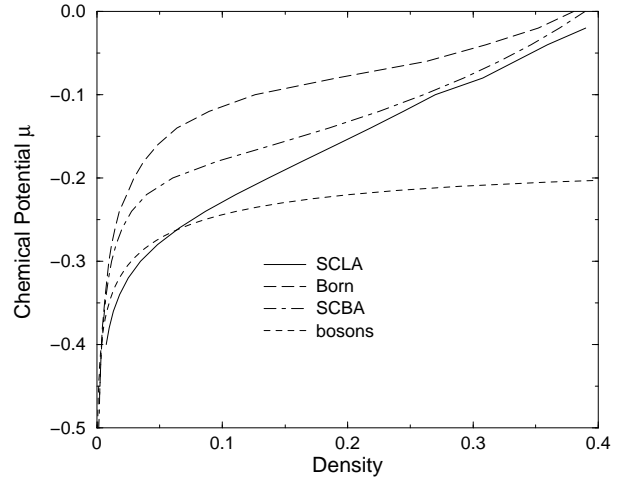,width= 8 cm}}  
\caption{Chemical potential $\mu$ in the three approximations, at
T=0.1.  The short-dashed line shows half the chemical potential of
non-interacting bosons, with ground state energy of $-E_b$.  }
\label{fig:mucomp}
\end{figure}

In Fig.\ref{fig:mucomp} we show the density dependence of the electron
(or hole) chemical potential as a function of the density for $T=0.1$,
in the three considered approximations. At very low density, $n \ll
0.01$, the chemical potential is $ \mu \ll - 0.4 $, and its value is
similar in all approximations. In this case, we are describing free
electron-hole pairs, as also suggested by a law of mass action, which
gives a crossover density of about $n^*=0.005$ at T=0.1. In this case
the description of $\mu$ is rather good even in approximations that
neglect existence of bound states. 
However, for $n > 0.01$, the
chemical potential in the Born approximations is much larger than for
the SCLA. In fact, in the SCLA we are also describing the fraction of
cold interacting excitons in the plasma.  As the ground state energy
of this part of the plasma is at about $-E_b$, the total energy of the
system is thus reduced with respect to that of a gas of unbound
particles. In order to clarify further this point, we plot on the same
graph half of the chemical potential of bosons at a ground state
energy of $-E_b$ and mass equal to the exciton mass. 
This is the chemical potential of a gas of electron (and
holes) completely bound into bosonic excitons. The halving comes from
the equilibrium condition $\mu_{X}=\mu_e+\mu_h=2 \mu$.  The chemical
potential of bosons compares reasonably well with the chemical
potential calculated in the SCLA, up to $\mu<-0.25$, $n<0.1$. Above
this limit, the chemical potential from the SCLA grows faster. There
are two reasons for this faster growth: first, the exciton gas is
repulsively interacting, second, the exciton gas eventually undergoes
a Mott transition and ionizes into free carriers. At densities
$n>0.2$, the SCLA and SCBA are indeed comparable, and at even larger
density, also the MBA approximation is reasonable. This shows that the
SCLA is a powerful tool for the investigation of the intermediate
regime of densities, where deviations from both the (non-interacting)
bosonic and pure fermionic models are important.

\section{Mapping the SCLA to a bosonic model at low density and temperature}
\label{sec:excpot}

\begin{figure} 
\centerline{\psfig{file=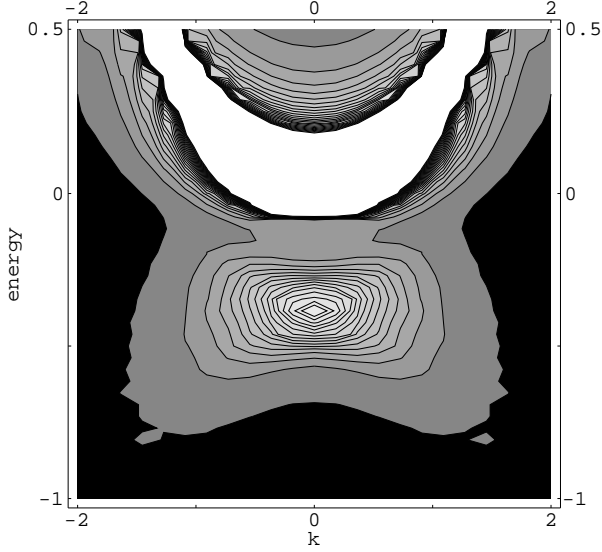,width= 8. cm}}
\caption{Contour plot of $A_e(k,\omega)$ at $T=0.04$ and n=0.02.}
\label{fig:spectral2}
\end{figure}

In Fig.\ref{fig:spectral2} we plot $A(k,\omega)$ for $T$=0.043 and
n=0.02. We can observe the parabolic dispersion of the main
quasi-particle peak, broadened by the scattering (white region in the
plot). Broadening is larger at small $k$ due to the larger phase space
in one dimension. For $\omega<0$, we have the correlation structure
shown for $k=0$ in Fig. \ref{fig:spectral1}, which is also present at
larger $k$.  We remark that this region of correlated electrons
extends into a region of $k$ of the order of 1 (i.e. of
$a_B^{-1}$). Its shape (dispersion) is related to the exciton
wavefunction, as we show in the following.

In a low-density, low-temperature limit, we can start to consider
$G=G_0$, and analytically calculate the retarded pair propagator
$H^+(k,\omega)$ from Eq. (\ref{eq:hr+}), neglecting $G_0^<$, which is
of the order of the density. We obtain:
$$ H^{(0)+}(k,\omega)\sim
\frac{-1}{\sqrt{2}\sqrt{\omega-\frac{k^2}{8}+2 i \gamma}},$$ where
$\gamma>0$ is the usual regularization number. Solving for the
T-matrix with the Bethe-Salpeter equation (\ref{eq:t+w}), and
expanding around its pole we obtain:
$$T^{(0)+}(k,\omega)=\frac{1}{\omega+\frac{1}{2}-\frac{k^2}{8}+2 i
\gamma}.$$ Thus, the T-matrix at the lowest order for the contact
potential has the form of a free propagator, for particles at an
energy $-1/2+k^2/8$, having a mass which is twice the electron mass,
i.e. the electron plus hole mass. Expanding the Bethe-Salpeter
equation to higher order, we obtain:
$$T^{(1)+}(k,\omega)=T^{(0)+}(k,\omega) H^{(1)+}(k,\omega)
T^{(0)+}(k,\omega).$$ Therefore, we may interpret $T^{(0)+}(k,\omega)$
as the free exciton (boson) Green function, and $ H^{(1)+}(k,\omega)$
as the lowest order self-energy.\cite{haussman} In particular
\begin{equation}\label{eq:hr1}
H^{(1)}(1,2)=2 i G^{(1)}(1,2) G_0(1,2),\end{equation}
$$G^{(1)}(k,\omega)= G_0(k,\omega) \Sigma^{(0)}(k,\omega)
G_0(k,\omega),$$
\begin{equation}\label{eq:sigmar0}
\Sigma^{(0)}(1,2)= i T^{(0)}(1,2) G_0(2,1).\end{equation} As
$T^{(0)}(k,\omega)$ is peaked close to $\omega=-1/2$, we may neglect
$G_0^<(k,\omega)$ in this region, and use
$$\Sigma^{(0)+}(k,\omega)\sim i\int\frac{dqd\omega'}{(2\pi)^2}
T^{(0)<}(q,\omega') G_0^{+}(k+q,\omega+\omega').$$ Integrating around
$\omega'=-1/2$, we obtain:
$$\Sigma^{(0)+}_k(\omega)\sim
\frac{n}{\omega+\frac{1}{2}+\frac{k^2}{8}+i\gamma},$$ where $n$ is the
total density. This is the structure shown in Fig. \ref{fig:self} at
$\omega=-1/2$, which produces the satellite correlation peak in the
electron spectral function, shown in Fig. \ref{fig:spectral2}. We also
notice that this structure has {\em negative} dispersion, as also
apparent in Fig. \ref{fig:spectral2}.

\begin{figure}
\centerline{\psfig{figure=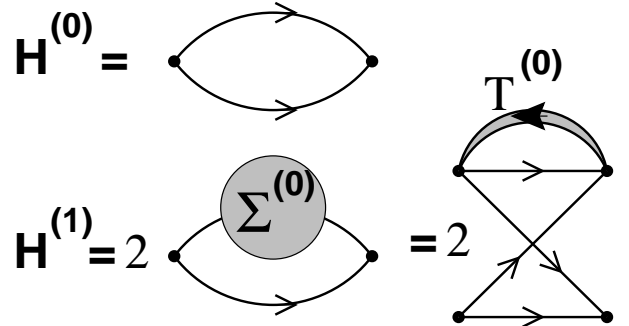,width=8truecm}}
\caption{The perturbation expansion to lowest order for the boson
(exciton) self-energy, and the four-particle vertex $F^{(0)}$.  Thin
lines are the bare electron (or hole) propagator. The shaded
two-particle Green function is the bare exciton propagator.}
\label{fig:excexc}\end{figure}

\begin{figure}
\centerline{\psfig{figure=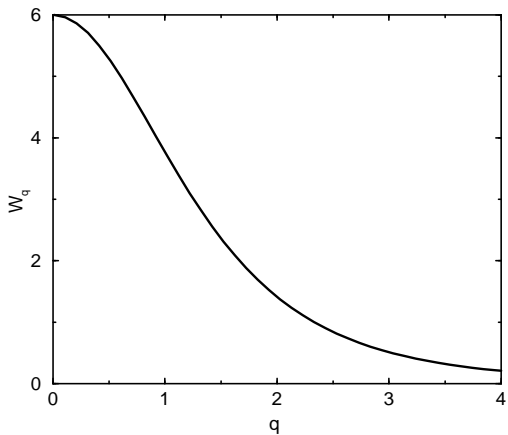,width=8truecm}}
\caption{ The exciton-exciton interaction potential $W_q=F^{(0)}(q,\omega_0)$.}
\label{fig:vq}
\end{figure}

We can now define an exciton-exciton interaction at low density, from
the Eqs. (\ref{eq:hr1}-\ref{eq:sigmar0}) above, which are pictorially
shown in Fig. \ref{fig:excexc}. We thus define a four-point
interaction kernel $F^{(0)}$, and the boson self-energy $H^{(1)}(1,2)$
is written as:
\begin{equation}
H^{(1)}(1,2)= \int d\bar{1}~ d\bar{2}~F^{(0)}(1,2;\bar{1},\bar{2})
T^{(0)}(\bar{2},\bar{1}). \end{equation} We can see from
Fig. \ref{fig:excexc} that $F^{(0)}$ represents the electron (or hole
exchange) in the scattering of the two excitons. Thus
\begin{equation}
 F^{(0)}(1,2;1'2') = 2 i G_0(1,2) G_0(1',2') G_0(1,2') G_0(1',2).
\end{equation} If we neglect retardation
(or memory) effects, we may define an instantaneous potential at the
exciton energy using $\tau_1=\tau_2$, $\tau_{1'}=\tau_{2'}$, and
calculating the resulting $F$ at the exciton frequency
$\omega_0=-1/2=E_B$. For simplicity, we also use $k,k'\sim 0$ for the
incoming excitons, and arrive to
$$F^{(0)}(q,\omega_0) \sim 2i \sum_{q',\omega'}
G_0^{+}(q',\frac{\omega_0}{2}+\omega')
G_0^{+}(-q',\frac{\omega_0}{2}-\omega')$$ \begin{equation}
G_0^{+}(q-q',\frac{\omega_0}{2}+\omega')
G_0^{+}(q'-q,\frac{\omega_0}{2}-\omega')=\end{equation}
$$=\frac{64}{q^2}\frac{2\sqrt{4+q^2}-(4-q^2)}{(4+q^2)^2}.$$ This
function rapidly drops to zero at $q>1$ as expected, and
$F^{(0)}(0,\omega_0)=6$.  This value is also obtained from standard
boson-boson exchange-interaction expressions (see
e.g. Ref. \onlinecite{haughanamura}) when only the electron-hole
interaction is used.  We show the full structure of
$F^{(0)}(q,\omega_0)$ In Fig. \ref{fig:vq}.

We are now in a position to discuss some of the qualitative changes
introduced by simplifying the Coulomb interaction to an on-site
one. In the limit of low density, we may in fact compare the
exciton-exciton interaction calculated above with that calculated
using a realistic Coulomb interaction, as a standard boson-boson
exchange expression at small momenta exists.\cite{tassone99xx} In a
realistic wire, a short-range cut-off is introduced by the finite size
of the electron and hole wavefunctions in the confinement directions.
A typical cut-off is of the order of the Bohr radius or
smaller. Typically, we have to consider tight confinements in order to
avoid participation of higher confined levels into the exciton
wavefunction. The Coulomb interaction is then reasonably represented
by a function $V_r=A/(r+r_0)$, where $r_0$ is a cut-off distance, and
$A$ is then chosen such that $E_b=-1/2$. 
We used $r_0=0.1$, and $A=0.18$. In this case the resulting Bohr radius
is 1, indicating tight confinement of the carriers in the confinement
plane. The exciton-exciton exchange interaction at small $q$ is
calculated in the small $q \ll {a_B}^{-1}$ limit
$$W_{xx}=2\sum_{k,k'}V_r(k-k')\phi_{1s}(k)\phi_{1s}(k')$$
\begin{equation}
\left[|\phi_{1s}(k)|^2-\phi_{1s}(k)\phi_{1s}(k')\right]\sim
1.4~, \label{eq:xxbose}
\end{equation}
where $\phi_{1s}(k')$ is the 1s wave-function of the exciton. The
positive term in Eq. \ref{eq:xxbose} is the exchange term due to the
attractive electron-hole interaction, while the negative term is due
to the electron-electron and hole-hole repulsive interactions.  In the
delta like-potential the negative term is absent because of
locality. In the long range case instead, the electron-electron (and
hole-hole) interaction largely cancel the electron-hole interaction in
the exchange integral, resulting in an exciton-exciton interaction
which is four times smaller when compared with the value of 6 obtained
for $F^{(0)}(q=0,\omega_0)$. We conclude that local interaction leads
to an over-estimation of the boson-boson interaction at a given
density, and therefore of broadening. However, this does not imply
that broadening effects are {\em qualitatively} different. Moreover,
we also stress that this overestimation does not concern at all the
comparison of the SCLA with the other simpler approximations
considered in this work, as all are carried out using the same
interaction potential.

\section{Optical Properties}
\label{sec:optical}

We now consider interaction of the electron-hole system with a
transverse electromagnetic field in the dipole approximation:
\begin{equation}
H_P=\sum_{{\bf q },q} C_{{\bf q},q} (a_{{\bf
q},q}P^{\dagger}_{q}+a^{\dagger}_{{\bf q},q}P_{q}),
\label{eq:Hp}
\end{equation}
where
$$ C_{{\bf q},q}=\frac{e p_{cv}}{m_0 c}\sqrt{\frac{4\pi \hbar c}{ q_t
V}} I_{\parallel}({\bf q}),
$$
is the dipole matrix element, which includes also the overlap integral
of the electromagnetic field with the confined carriers
$I_{\parallel}$.  Here ${\bf q }$ represents 2D wavevectors in the
plane orthogonal to the wire axis, $q$ are wavevectors along the wire,
while $q_t^2={\bf q}^2+q^2$, $m_0$ is the free electron mass, $c$ the
velocity of light, and $p_{cv}$ the momentum matrix element between
the conduction and hole bands.  $a_{{\bf q},q}$ is the photon
destruction operator, and $P_{q}$ is the polarization operator,
defined as $P_{q}=\sum_{k}d_{q-k}c_{k}$, i.e. local in real space in
the dipole approximation.  As the section of the wires is always much
smaller than the wavelength of light, we have $I_\parallel\sim
1$. Moreover $q_t \sim E_{GAP}/(\hbar c)$ is almost constant, thus we
use arbitrary units from now on, with $C=1$. Actual values of
absorption and gain are easily calculated by using the appropriate
value of $C$ for the given material.

In the process of scattering of light from the system, an absorbed
photon creates a coherent electron-hole pair which propagates in the
crystal, interacts with the background plasma, then recombines, and a
photon is emitted.  When we introduce photon propagators, the
interaction process is represented by a photon self-energy. We will
not solve the photon Dyson equation, nor we will dress the electron
and hole propagation with the photon interaction. In other words, we
neglect both polaritonic effects, and higher order non-linear
interactions. In fact, we are restricting our attention to weak
external fields, i.e. assuming interaction of the plasma with the
photon field to be weak. Otherwise, it could be possible for the
system to be driven out of equilibrium, because the photon chemical
potential is zero and far below that of the elecron-hole plasma. This
equilibrium condition is well fulfilled in real systems, exception
made for the strongest laser fields.  Within this weak interaction
approximation, the photon self-energy is exactly given by the pair
correlation function
\begin{equation}
\Pi_q(\tau,\tau')= \langle T P^{\dagger}_{q}(\tau)
P_{q}(\tau')\rangle.\label{eq:pairprop}\end{equation} However, this
correlation function has to be calculated within some approximation,
and also using approximate propagators.  The resulting self-energy may
not be conserving. In particular, particle flux conservation at the
vertex translates in the longitudinal f-sum rule\cite{baym61}, which
has to be respected in any reasonable approximation.

In order to shed light on this important issue, we consider the
problem of interaction of photons with the plasma from a different
viewpoint. As we are only interested in the linear response to the
external electromagnetic field, the correct photon self-energy is also
given by the linear response of the plasma to a {\em classical}
electromagnetic field, i.e. considering the photon creation and
destruction operators in Eq. (\ref{eq:Hp}) as c-numbers.  Baym and
Kadanoff give a recipe to construct a conserving expression for this
response function in Ref.\onlinecite{baym61}. When an external
coherent electromagnetic field is applied to the plasma, coherence
develops in the plasma, and pair propagators have additionally to be
defined along with the single particle propagators. They are analogous
to anomalous propagators in the theory of superconductivity. The
electron-hole anomalous propagator reads
\begin{equation}\label{eq:pairprop2}
G_{eh}(1,2)=i\langle T d(2) c(1) \rangle. \end{equation} 
Times are defined on the Keldysh contour as usual. Another anomalous
propagator $G_{he}$ can be similarly defined for convenience, and
notation regrouped into a matrix one, like for Nambu propagators in
the problem of superconductivity.\cite{nambu60} Dyson
Eq. (\ref{eq:dyson}) is extended to include both 
anomalous propagators, and the external interaction $H_p$:
$$
\int d\bar{2}~G_0^{-1}(1,\bar{2})G(\bar{2},3)=\delta(1,3)+\int
d\bar{4}~\Sigma(1,\bar{4})G(\bar{4},3)+$$
\begin{equation}
+\int
d\bar{4}~\Sigma_{eh}(1,\bar{4})G_{he}(\bar{4},3)-A(1)G_{he}(1,3)\end{equation}
$$\int d\bar{2}~G_0^{-1}(1,\bar{2})G_{eh}(\bar{2},3)=\int
d\bar{4}~\Sigma(1,\bar{4})G_{eh}(\bar{4},3)+$$
\begin{equation}+\int d\bar{4}~\Sigma_{eh}(1,\bar{4})G(\bar{4},3)- A(1)G(3,1)
\label{eq:dysonanomal}\end{equation}

As a general rule for generating a conserving approximation, the self
energies have to be obtained from functional derivation of a
grand-potential with respect to the propagators. The normal
self-energies have been already introduced.  For the anomalous self
energy $\Sigma_{eh}(1,2)$, we need to generalize the expression of the
grand-potential including terms where the normal propagators are
replaced by the anomalous ones.  The anomalous propagators $G_{eh}$
are at least first order in the external potential $A$, as spontaneous
symmetry breaking is ruled out by the Mermin-Wagner theorem.  For
linear response, we need the anomalous self energy $\Sigma_{eh}$ up to
linear terms in the external potential only, thus at most linear terms
in the anomalous propagators. Consequently, we need at most quadratic
terms in these propagators in the grand-potential to obtain the
$\Sigma_{eh}$ from the functional derivative. We conclude that the
only and sufficient term to be considered is the anomalous Fock term,
which is the usual Fock term with anomalous propagators replacing
normal ones. As the approximation is conserving, we are also
guaranteed that absorption fulfills the f-sum rule\cite{baym61} for
any value of density and temperature.  The anomalous self-energy
obtained from the anomalous Fock term reads
\begin{equation}\Sigma_{eh}(1,2)=-i G_{eh} (1,2)+O(A^2).\label{eq:sigmaanomal}
\end{equation}
Normal $G$ are at least second order in the external potential $A$,
while both $G_{eh}$ and $\Sigma_{eh}$ are at least first order. The
anomalous Dyson Eq. (\ref{eq:dysonanomal}) to lowest order becomes:
$$\int d\bar{2}~G^{-1}(1,\bar{2})G_{eh}(\bar{2},3)=-i \int
d\bar{4}~G_{eh}(1,\bar{4})G(\bar{4},3)-$$
\begin{equation}
- A(1) G(1,3).
\label{eq:dysonanomal2}\end{equation}
Introducing the response function
\begin{equation}
\tilde \Pi (1,2)= -i \left.\frac{\delta
G_{eh}(1,1)}{A(2)}\right|_{A=0},
\end{equation}
and taking the first order variation of Eq. (\ref{eq:dysonanomal2}),
we obtain and equation for $\tilde \Pi (1,2)$:
$$
\tilde \Pi (1,2)= i G(1,2)G(1,2) -i\int d\bar{3}~ G(1,\bar{3})
G(1,\bar{3}) \tilde \Pi (\bar{3},2)=$$
\begin{equation}=
H(1;2) - \int d\bar{3}~ H(1;\bar{3}) \tilde \Pi (\bar{3},2),
\end{equation}
which is easily solved and gives
\begin{equation} 
\tilde \Pi (1,2)= H(1;2) +\int d\bar{3}~ d\bar{4}~ H(1;\bar{3}) T
(\bar{3};\bar{4}) H (\bar{4};2). \end{equation} It is straightforward
to show that $\tilde \Pi (1,2)$ obeys bosonic Kubo-Martin-Schwinger
relations, as both $H$ and $T$ do. In particular, we have:
\begin{eqnarray} 
\tilde \Pi^+(q,\omega)= H^+(q,\omega)\left|  T^+(q,\omega)\right|^2 \\ 
\tilde \Pi^<(q,\omega)= H^<(q,\omega)\left|  T^+(q,\omega)\right|^2.
\end{eqnarray}
Absorption $\alpha(\omega)$ and photoluminescence $PL(\omega)$ at
$q=0$ are derived as $\alpha(\omega)=-\Im [\Pi^+(0,\omega)]$ and
$PL(\omega)=-\Im [\Pi^<(0,\omega)]$ respectively.

We remark that higher order terms than the anomalous Fock one have to
be considered in the anomalous self-energy expansion when a {\em
finite} external field is present, such as in the case of a strong
laser field.  Anomalous Born terms calculated in the Markov
approximation have been considered in the
semiconductor Bloch equations by Lindberg and Koch in
Ref. \onlinecite{lindberg}, and named polarization-polarization
scattering terms. Here we do not further pursue such extensions, which
are clearly beyond the scope of the present paper. We only remark that
our approach is easily extended to such conditions, while keeping full
control of the conservation laws.

\subsection{Optical spectra and stability of the excitonic emission}

\begin{figure}
\centerline{\psfig{file=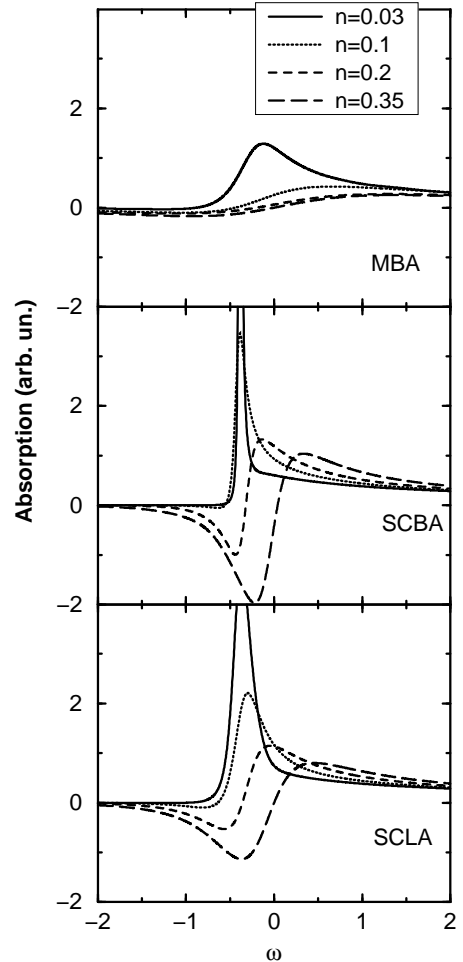,width=11 truecm}}
\caption{Absorption spectra in the Markof-Born Approximation (MBA),
self consistent Born approximation (SCBA) and self consistent ladder
approimation (SCLA) for different densities indicated in the figure.}
\label{fig:abs}
\end{figure}

We show in Fig. \ref{fig:abs}, the absorption spectra in the normal
direction ($q=0$) at $T=0.2$, for different densities, in the MBA,
SCBA, and SCLA. In the absorption spectrum, we remark at low density
the characteristic excitonic peak and the continuum absorption. The
exciton linewidth is extremely narrow in the SCBA, and much broader in
the MBA. The SCLA linewidth is intermediate. In the SCBA there is a
very small broadening at the exciton energy, as no excitons are
represented in the theory, while in the MBA, the free carrier
broadening at $\omega=0$ is assumed. This is somewhat larger than the
broadening in the SCLA at $T=0.2$.  For large density $ n>0.2$, the
excitonic peak is bleached in all approximations, and a region of
negative absorption, i.e. optical gain, appears. The absorption
changes sign at $\omega=\mu_e+\mu_h=2\mu$, coinciding with the change
of sign of the Bose factor, and resulting in a well defined
-i.e. positive- photoluminescence through the Kubo-Martin-Schwinger
relation.

\begin{figure} 
\centerline{\psfig{file=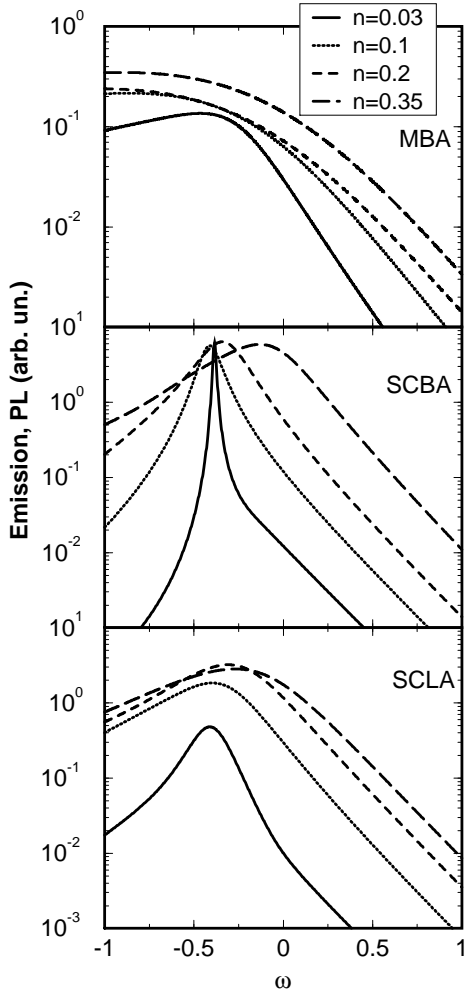,width=11 cm}}
\caption{Photoluminescence spectra in the MBA, SCBA and SCLA for different
densities indicated in the figure.}
\label{fig:pl}
\end{figure}

We plot the emission or PL spectra in Fig. \ref{fig:pl}. First we
notice the different scale for the MBA, where a smaller peak emission
is calculated even for the largest considered density.  For large
density $n>0.2$, we observe a saturation of the intensity. This can be
understood in terms of the fermionic nature of the carriers, when the
electron-hole plasma becomes degenerate. Second, we notice
exciton-like emission even when excitonic absorption is bleached in
all models.  In fact, vertex corrections place the free carrier
emission at the exciton energy, even when no bound excitons are
described in the gas. 
Eventually, only at very small density the free
carrier and excitonic emission become distinguishable, as we notice an
exponential emission shoulder at high energy $\omega>0$ in the SCLA
and SCBA, reminding of a fermionic emission tail at small degeneracy
(Boltzmann distribution). At larger density, the two features at
different energy merge into a unique one, due to increased
broadening. A single peak is also observed in experiments, where
finite noise in the data and large inhomogeneous broadening mask any
minor feature.\cite{optwires} For this reason, it is in practice very
delicate to establish the position of the band-gap (bottom of the
free-carrier bands) directly from emission data.\cite{piermarocchi99}
Third, we remark that the low-energy shoulder of the PL emission in
the MBA is Lorentzian because of the Markov approximation, and the
broadening largely over-estimated also at low energies. The linewidth
in the SCBA is instead underestimated at low-density, as for
absorption.

\begin{figure} 
\centerline{\psfig{file=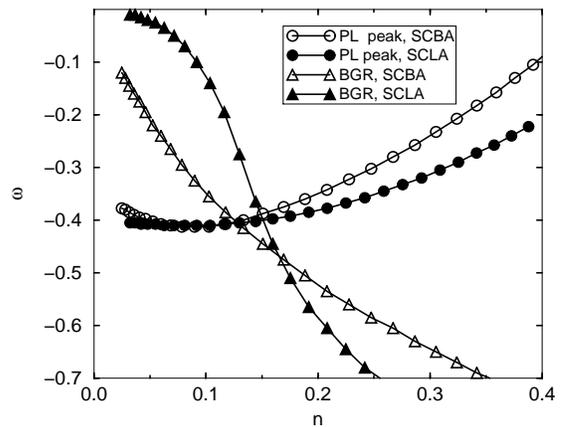,width=8 cm}}
\caption{The band gap renormalization (BGR) and emission peak energy
in the SCBA and SCLA as indicated in the figure.}
\label{fig:stability}
\end{figure}

 An interesting -and directly observed- physical quantity, is the
energy of the PL peak. As remarked above, emission is excitonic due to
the correlation of the coherent electron-hole pair emitted. In
Fig. \ref{fig:stability} we plot the position of this peak in the PL
as a function of the carrier density, together with the band-gap
renormalization (BGR) defined as twice the energy of the main peak of
the spectral function. Interestingly, exciton bleaching in absorption
appears at a density which is comparable to that where the band gap
renormalization crosses the emission energy, i.e. $n\sim 0.15$. This
can therefore be assigned as a Mott density, i.e. the energy where the
binding energy becomes negligible with respect to broadening, thus
effectively vanishing. There are two interesting features to be
observed in Fig. \ref{fig:stability}: first, the band gap
renormalization is negligible in the SCLA for $n<0.1$, second,
emission energy is constant in this range of densities, and
blue-shifting less than in the SCBA for larger density. Comparison
with the MBA is vitiated by excessive broadening in this
approximation. The simple Hartree-Fock approximation instead gives
results which are similar to those obtained in the SCBA for the band
gap renormalization and the PL emission peak, apart from more
pronounced blue-shifts.\cite{tassone99} Stability of the emission peak
is usually interpreted as a partial compensation of the self-energy
and vertex corrections. As in the Hartree-Fock approximation the
broadening effects are missing, we deduce that these effects are
indeed relevant for the cancellation found in the SCBA and in the SCLA
at high density, and explain the reduction of the blue-shift with
respect to simple Hartree-Fock calculations.

\subsection{Excitonic gain}

Optical gain is usually considered in the regime of degenerate
electron-hole plasma. In this regime, spectral functions can be
assumed to be simple Lorentzians, and the absorption can be written
as:
\begin{equation}\label{eq:gainfc}
\alpha(\omega)\propto\int \frac{dk}{2 \pi} \frac{1-2 f(k^2/4
-\mu)}{(\omega-k^2/2)^2+ \gamma^2},
\end{equation}
where $\gamma$ is the spectral broadening. Then, negative absorption
or gain occurs only when the Pauli blocking factor ($1-2 f(k^2/4
-\mu)$) becomes negative, which necessarily requires a chemical
potential above the band gap, or in other words, inversion. However,
both vertex corrections, and deviations of the spectral function from
simple Lorentzian should be taken into account at lower density and
temperature. In this case,
absorption can be written as:
\begin{eqnarray}
\alpha(\omega) \propto |T^{+}(q=0,\omega)|^2
\int\frac{dk}{2\pi}\frac{d\omega^\prime}{2 \pi}\times
\label{eq:abstherm}
\end{eqnarray}$$
\left[1-f(\omega-\omega^\prime-\mu)-f(\omega^\prime-\mu)\right]
A_{k}(\omega-\omega^\prime)A_{k}(\omega^\prime)~,
$$
The term in square brackets
$1-f(\omega-\omega^\prime-\mu)-f(\omega^\prime-\mu)$ is the
generalization of the Pauli blocking factor for a system where the
quasiparticles are described by arbitrary spectral functions, and
clearly one of the Fermi functions refers to electron occupation, the
other to hole occupation. This expression of the absorption is valid
only in the case of short range potential, and for the long range case
$T^+$ depends also on the relative momentum of the pair $k$, and
becomes part of the integral kernel.  Using the definition of Fermi
and Bose functions, the absorption can be recast in the form
\begin{eqnarray}
\alpha(\omega)=\frac{|T^{+}(q=0,\omega)|^2}{g(\omega-2 \mu)}
\int\frac{dk}{2\pi}\frac{d\omega^\prime}{2\pi} \times\label{eq:absg}
\end{eqnarray}$$\times
f(\omega-\omega^\prime-\mu)f(\omega^\prime-\mu)
A_{k}(\omega-\omega^\prime)A_{k}(\omega^\prime)~,$$ where
$g(\omega-2\mu)$ is the Bose function.  Therefore, gain clearly occurs
below twice the chemical potential $\mu$ (i.e. $\mu_e+\mu_h$), even
when the chemical potential is {\em below} the band-gap. However, for
it to be sizeable, the spectral functions must have non-negligible
weight in this region of the spectrum.  In the MBA and SCBA case, this
weight is clearly given by broadening effects alone, while in the SCLA
it is also due to the presence of the excitonic correlation peak in
the spectral function as shown in Fig.\ref{fig:spectral1}.  A sizeable
enhancement of the gain is found in correspondence to the excitonic
resonance because of the vertex correction, given by the $|T^+|^2$
factor in Eq. (\ref{eq:absg}). When the $2\mu< -E_B$, both gain and
excitonic absorption coexist. In this case, we talk of {\it excitonic
gain}. The system is inversion-less in the sense of
Eq. (\ref{eq:gainfc}) above. However, only in the SCLA we are
describing gain due to the presence and scattering of excitons in the
low temperature, low-density plasma, while in the other
approximations, gain is purely related to  dynamical effects in the
interaction vertex with the photon.

\begin{figure} 
\centerline{\psfig{file=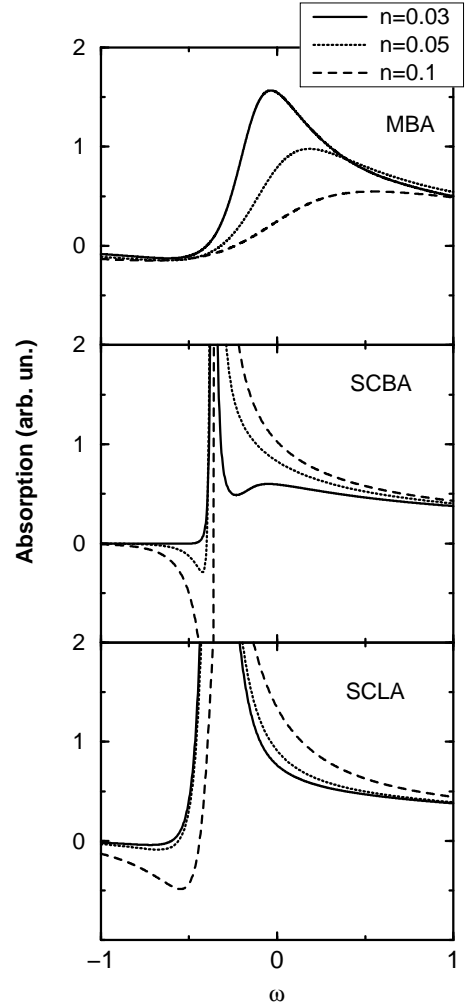,width=11 cm}}
\caption{Absorption spectra at $T=0.1$ for the MBA, SCBA, and SCLA,
and low densities indicated in the figure.}
\label{fig:gain}
\end{figure}

In Fig.\ref{fig:gain} we plot the absorption spectra close to the
exciton resonance at $T=0.1\ll E_B$. For the considered densities
$n<0.1$, gain coexists with the excitonic resonance in absorption.
Only for the MBA at $n=0.1$, the absorption peak is completely shifted
to higher energies, and we can not anymore speak of excitonic
absorption. As usual, the broadening in the exciton spectral region is
too small and unphysical in the SCBA, resulting in very sharp and
unphysical features. Gain in the SCLA is instead over a larger
spectral region, and its value is larger than what predicted in the
MBA at large density $n>0.3$ shown in Fig.\ref{fig:abs}. Excitonic
gain clearly originates from inclusion of excitons in the plasma for
the SCLA. In fact, reasonable broadening is calculated in the exciton
spectral region, and mainly originating from exciton-exciton
scattering as shown in Sec. \ref{sec:excpot}. We conclude, that the
SCLA indicates that sizeable excitonic gain can be obtained at
moderate density and temperature, in a reasonably large spectral
region of the order of a fraction of $E_B$.

Excitonic gain has been widely studied in II-VI quantum wells
\cite{kozlov96}.  However these are two-dimensional system, and
excitonic correlations are expected to be less pronounced in this
systems, so at this stage it is not reasonable to make even a
qualitative comparison.  For quantum wires, sizeable gain at 10 K
($T\sim 0.05$) for an estimated carrier density well below the Mott
has been recently claimed by Sirigu {\it et al}.\cite{sirigu99} This
is a first indication that excitonic gain might be relevant in these
systems. However, a qualitative and quantitative understanding of this
experiment is clearly premature. From the experimental side,
inhomogeneous broadening due to interface disorder has to be decreased
well below the binding energy, while a more realistic Coulomb
interaction has to be addressed by the theory.

\section{Conclusions}
\label{sec:conclusions}

We have presented a model that includes excitonic correlation in the
description of a highly-excited semiconductor quantum wire. The model
has been simplified using a short range potential, and considering a
polarized gas.  Correlation has been calculated self-consistently at
the ladder level (SCLA). We have shown that bound states appear as a
low-energy correlation peak in the spectral function of electrons and
holes.  We compared the results obtained with SCLA to the ones
obtained within lower order approximations that do not include
excitonic correlation (Born approximations).  Even though the model is
purely fermionic, we have shown how it can be effectively mapped at
low temperatures and density to a gas of interacting excitons. We have
thus derived an analytical expression for the effective
exciton-exciton interaction.  The linear optical properties of the
system have been calculated including vertex corrections at the Fock
level. This ensures the conservation of sum rules in the optical
response. The excitonic absorption at low density ensues both in
ladder and Born approaches, but the broadenings at the exciton energy
are either too small or too large in the Born approximations,
depending on whether frequency dependence of the broadening is
included or not. Excitonic emission well beyond exciton bleaching is
also predicted in all models, but the peak shifts are more pronounced
in the Born approximations, showing that a better cancellation between
the self-energy and vertex correction results from the introduction of
excitons. We have also shown that sizeable excitonic gain can be
predicted at low temperature and density, when the electron plus hole
chemical potential is just below the exciton energy. However, its
correct description must include exciton broadening from
exciton-exciton scattering, and unphysical values are thus obtained in
the Born approximations. These qualitative conclusions clearly hold
even for more refined descriptions of the electron-hole plasma than
the Born approximations considered here, when these descriptions do
not account for excitons in the plasma. For example this is the case
for the model recently presented by Das Sarma and Wang in
Ref. \onlinecite{dassarma00}, which includes screening at the
plasmon-pole approximation level, but no excitons in the electron-hole
gas. The excitonic gain calculated using this approximation shows a
divergent behavior close to $-E_b$, similar to what we have found for
the SCBA. The extension of the proposed approach to a full
non-equilbrium theory could contribute to a deeper understanding of
many body effects in electron-hole systems, also in the perspective of
ultrafast optical experiments than can investigate far-from
equilibrium conditions. In this context a recent theory by Hannewald
{\it et al.}\cite{hannewald00} provides a convincing step in this
direction, yet remaining within a non-dynamical approach.  The last
issue concern the intrinsic limitations of the SCLA. It has been shown
that in order to describe correctly the very diluite excitonic limit
additional diagrams must be added to the self consistent ladder
approach. An analysis of this problem can be found in
Ref.\onlinecite{pieri98}, where a generalized T matrix approximation
for the fermionic self energy overcoming this problem has ben
proposed.  In conclusion, we have clearly shown that in the definite
and important physical region of $T< E_b$ inclusion of excitonic
correlation in the electron-hole plasma is relevant and necessary, and
that the simplified model presented in this paper is a well-understood
starting point for this purpose. Thus, its further developments to
address realistic systems in higher dimensions are well motivated.

\acknowledgments We thank A. Quattropani, P. Schwendimann, V. Savona,
C. Ciuti, and L. J. Sham for stimulating discussions. One of the
authors (C. P.) acknowledges support by the Swiss National Foundation
for the Scientific Research.

\end{document}